\documentclass[fleqn,usenatbib]{mnras}

\usepackage{newtxtext,newtxmath}
\usepackage{hyperref}
\usepackage{multirow} 
\usepackage{tabularx}
\usepackage{xcolor} 
\usepackage{bm}

\usepackage[T1]{fontenc}

\DeclareRobustCommand{\VAN}[3]{#2} 
\let\VANthebibliography\thebibliography
\def\thebibliography{\DeclareRobustCommand{\VAN}[3]{##3}\VANthebibliography}

\usepackage{graphicx}	
\usepackage{amsmath}	

\newcommand{\revone}[1]{#1} 
\newcommand{\revtwo}[1]{#1} 

\newcommand{\breifast}{\textsc{Breifast}}
\newcommand{\stimela}{\textsc{Stimela2}}

\newcommand{\xarray}{\mbox{\textsc{XArray}}}
\newcommand{\xarrayfits}{\mbox{\textsc{XArray-FITS}}}
\newcommand{\pybdsf}{\mbox{\textsc{PyBDSF}}}
\newcommand{\oxkat}{\mbox{\textsc{OxKAT}}}

\title[TRON I. MSPs in globular clusters]{Mining the time axis with TRON. I. Millisecond pulsars in Omega Centauri, Terzan 5 and
47 Tucanae detected through MeerKAT interferometric imaging}

\author[Smirnov, O.M. et al.]
{\parbox{\textwidth}{
\begin{flushleft}
O.~M.~Smirnov$^{1,2,3}$\thanks{E-mail: o.smirnov@ru.ac.za},
I.~Heywood$^{4,7,1,2}$, 
M.~Geyer$^{5}$, 
T.~Myburgh$^{1}$, 
C.~Tasse$^{6,1}$, 
J.~S.~Kenyon$^{1}$, 
S.~J.~Perkins$^{2}$,
J.~Dawson$^{1,2}$, 
H.~L.~Bester$^{2,1}$, 
J.~S.~Bright$^{4,7}$, 
B.~Ngcebetsha$^{2,1}$, 
N.~Oozeer$^{2,1}$,
V. G. G.~Samboco$^{1}$,
I.~Sihlangu$^{2,1}$, 
C.~Choza$^{4,7}$,
A.~P.~V.~Siemion$^{7,4,8,9,10}$
\\
\end{flushleft}
}
\\
\footnotesize
\\
$^{1}$Centre for Radio Astronomy Techniques and Technologies (RATT), Department of Physics and Electronics, Rhodes University, Makhanda, 6140,\\South Africa\\
$^{2}$South African Radio Astronomy Observatory, Cape Town, 7925, South Africa\\
$^{3}$Institute for Radioastronomy, National Institute of Astrophysics (INAF IRA), Via Gobetti 101, 40129 Bologna, Italy \\
$^{4}$Astrophysics, Department of Physics, University of Oxford, Keble Road, Oxford, OX1 3RH, UK\\ 
$^{5}$High Energy Physics, Cosmology \&  Astrophysics Theory (HEPCAT) Group, Department of Mathematics and Applied Mathematics, \\ \,\,University of Cape Town,  Rondebosch 7701, South Africa\\
$^{6}$GEPI \& ORN, Observatoire de Paris, Université PSL, CNRS, 5 Place Jules Janssen, 92190 Meudon, France\\
$^{7}$Breakthrough Listen, Astrophysics, Department of Physics, University of Oxford, Keble Road, Oxford, OX1 3RH, UK\\
$^{8}$SETI Institute, 339 Bernardo Ave, Suite 200, Mountain View, CA 94043, USA\\
$^{9}$Department of Physics and Astronomy, University of Manchester, UK\\
$^{10}$University of Malta, Institute of Space Sciences and Astronomy, Msida, MSD2080, Malta
}

\date{Accepted 2025 January 17. Received 2025 January 17; in original form 2024 October 29}

\pubyear{2024}

\begin{document}
\label{firstpage}
\pagerange{\pageref{firstpage}--\pageref{lastpage}}
\maketitle

\begin{abstract}

Medium-timescale (minutes to hours) radio transients are a relatively unexplored population. The wide field-of-view and high instantaneous sensitivity of instruments such as MeerKAT provides an opportunity to probe this class of sources, using image-plane detection techniques. We aim to systematically mine archival synthesis imaging data in order to search for medium-timescale transients and variables that are not detected by conventional long-track image synthesis techniques. We deploy a prototype blind transient and variable search pipeline named TRON. This processes calibrated visibility data, constructs high-time cadence images, performs a search for variability on multiple timescales, and extracts lightcurves for detected sources. As proof of concept, we apply it to three MeerKAT observations of globular clusters, known to host transient or variable sources. We detect a previously known eclipsing MSP suspected to be a `black widow' system, in the globular cluster Omega Centauri, with a light curve confirming the eclipsing nature of the emission. We detect a previously known `red back' eclipsing MSP in the globular cluster Terzan 5. Using observations of the globular cluster 47 Tucanae, we detect two known millisecond pulsars (MSPs), and one previously reported MSP candidate, with hints of eclipsing behaviour. 
\end{abstract}

%

\section{Introduction}

MeerKAT's high instantaneous sensitivity and large field of view make it an excellent instrument for detecting radio transients.
The ThunderKAT large science program (LSP) has yielded a wide range of important discoveries \citep[see e.g.][]{thk-andersson,andersson-citizen-science,thk-driessen1,thk-driessen2}. Typically, these have been either triggered by dedicated transient search engines such as MeerTRAP \citep{meertrap1,meertrap2}, or have involved targeted multi-epoch observations of particular fields. \revtwo{Recent results \citep[e.g.][]{chastain2023,dobie2023,wang2023,deruiter2024,fijma2024} suggest that more routine imaging observations can also yield a rich vein of medium-timescale (minute-to-hour) radio transients or variables. \revtwo{For example, \citet{parrot}} report on the serendipitous discovery of the PARROT, a pulsar-type object, during observations of the Great Saturn--Jupiter Conjunction of 2020, \revtwo{while \citet{47Tuc}} reports a millisecond pulsar (MSP) candidate detected in MeerKAT imaging observations of the globular cluster 47 Tucanae (47 Tuc)}. In both cases,  medium-timescale spectro-temporal variability (whether intrinsic, or caused by refractive effects along the line-of-sight) is the observational signature. With almost six years' worth of MeerKAT imaging observations already in the archive, a dedicated program of mining this data for more such discoveries seems well worth the effort. 

Such an effort can also have important implications for the search for technosignatures. For a variety of intrinsic and extrinsic reasons, many, perhaps most, radio technosignatures are expected to be variable and thus would form a subset of the population of astronomical transient events. Current technosignature surveys \citep[e.g.][]{wordenLISTEN} typically use a highly targeted approach, with either single dish or beamformed observations of individual sources and specific search algorithms for a limited set of spectrotemporal morphologies.  This is the approach currently in-use by the Breakthrough Listen User Supplied Equipment commensal program on MeerKAT \citep{czechmeerkat}. However, given that an interferometer samples the entire field of view simultaneously, and that with arrays formed by large numbers of small dishes the field-of-view can be relatively large, a systematic transient mining program in the image plane offers a complementary and more generic approach.

\revone{In this series of letters, we report on initial results from a prototype pipeline we have deployed for mining image-plane transients from radio interferometric imaging data.}

\section{Observations and data reduction}

The observations reported on here were conducted by MeerKAT in conventional synthesis imaging mode, using the L-band (856--1712 MHz) system in 4096 channel correlator mode, with 8s integrations. They consisted of multiple-hour tracks of the target, interspersed with shorter scans on a bandpass and gain calibrator.

47 Tuc was observed on 17 June 2018, using J0252$-$7104 as the secondary gain calibrator. The total on-target time was 5.17~h. Observations of $\omega$ Cen took place on 11 November 2018, using J1318$-$4620 as the secondary, for a total on-source time of 9.13~h. J0408$-$6545 was used as the primary bandpass calibrator in both cases. Terzan 5 was observed on 6 June 2019, with J1939$-$6342 and J1830$-$3602 used as the primary and secondary calibrators respectively. The observations took place during MeerKAT's science verification phase, thus the calibration strategy was conservative, with frequent visits to calibrator sources. This is reflected in the temporal gaps that are visible in the results that follow.

\paragraph*{Data reduction.} Our data processing pipeline requires a conventional reference calibration procedure, followed by imaging and self-calibration, followed by an optional direction-dependent (DD) calibration and peeling step (not necessary for the data at hand). This is followed by a separate detection-and-analysis stage called TRON (Transient Radio Observations for Newbies), implemented as a \stimela\ \revtwo{\citep{stimela2}} recipe. The inputs to TRON consist of (a) a best-effort \revone{(see below)} deep image of the target field, usually obtained via multi-frequency synthesis (MFS) imaging of the full-track data, and (b) residual visibilities, computed as calibrated visibilities minus model visibilities corresponding to the deep image. 

TRON itself consists of the following stages:

\begin{itemize}
   \item HTC imaging of the residual visibilities \revone{as per \citet{parrot}}. A separate snapshot image is produced per each integration in the measurement set (in this case, every 8s). \revone{No deconvolution is done on the snapshots}.
   \item Stacking of the images into a time cube, stored as an \xarray\ dataset using the \xarrayfits\ package\footnote{\url{https://github.com/ratt-ru/xarray-fits}}
   \item \revone{Smoothing the raw time resolution cube to a set of longer timescales so as to increase sensitivity to longer-duration events. \revtwo{The set of timescales is a tunable parameter, with the default setting of 15~s to 960~s in successive increments of $\times2$ used in this work.}}
   \item Detection of sources in the deep MFS image using the \pybdsf\ source finder \citep{pybdsf}.
   \item Detection of candidates using \breifast, a new detection algorithm forming part of TRON. This consists of peak finding in the per-timestamp images, followed by a number of heuristics to weed out the many false positives induced by primary beam rotation, as well as residual atmospheric and instrumental effects not handled by calibration, and remaining weak radio frequency interference (RFI). These effects tend to average out in long synthesis images (conversely, effects that \emph{do not} average out would be handled by normal self-calibration in the first place), but can be quite prominent in the per-integration images.
   
   Since \breifast\ operates purely in terms of residual image cubes, it will report both substantial variability of existing sources picked up by \pybdsf, as well as ``pure'' transients that would not be detected in the deep image due to their relatively short duration. 

   \revtwo{For this work, we used the current default \breifast\ settings of a $7\sigma$ peak threshold (decreasing to $6\sigma$ at the longest two timescales) with respect to a local r.m.s. estimate, and a minimum 10\% excursion relative to the baseline flux measured in the deep image. This served to exclude any false-positives that would otherwise be induced by instrumental effects such as primary beam rotation and varying $uv$-coverage.}

   \revone{Once the detections are made, TRON proceeds to extract data products for these (and/or any other sources deemed interesting by the user), using the methodology described by \citet{parrot}. The results in Sect.~3 are based on these products:}
   
   \item Light curves towards sources of interest.
   \item Dynamic spectra, extracted using the {\sc DynSpecMS} tool\footnote{\url{https://github.com/cyriltasse/DynSpecMS}}. This tool performs the synthesis of dynamic spectra from inteferometric data at the native time and frequency resolution, \revtwo{and is discussed in detail by \citet{Tasse2025}.} We have previously used this tool with LOFAR and MeerKAT data \citep{CRDraconis,parrot}, with the latter paper also providing a brief description.
\end{itemize}

Since the inputs to TRON are, generally, yielded by any conventional calibration and imaging workflow, it is largely agnostic to \emph{what} is used for the calibration stage, but highly sensitive to the \emph{quality} of that stage. \revone{Detectability is ultimately limited by the r.m.s. level of the snapshot images, which is in turn driven by the completeness of the subtracted model (i.e. the deep image), as well as any remaining undeconvolved flux, calibration artefacts, and unflagged low-level RFI. See Sect.~4 of \citet{parrot} for a discussion.
As part of our mining effort, we are working on a more holistic calibration pipeline optimized for transient detection, which (along with the full details of TRON) will be discussed in a follow-up technical paper.}

The observations of 47 Tuc and $\omega$ Cen reported on here were reduced by the \oxkat\ pipeline \citep{heywood2020}, following the process described by \citet{heywood2022}. 
\revone{For deep imaging, we produced $10125\times10125$ images with a pixel size of $1\farcs1$, using robust=0 weighting, reaching a noise level of 3.1 and 3.5 $\mu\mathrm{Jy~beam}^{-1}$ for 47 Tuc and $\omega$ Cen respectively. This resulted in a snapshot image noise level of 120--170 $\mu\mathrm{Jy~beam}^{-1}$.}

The Terzan 5 data were calibrated manually, as the scans of the primary were corrupted, and as such we used the scans of the secondary calibrator to derive all gain solutions. These were solved for against a model retrieved from the ATCA Calibrator Database\footnote{\url{https://www.narrabri.atnf.csiro.au/calibrators/calibrator_database.html}}. Fortunately, this calibrator is bright (8.47 Jy at 1.28~GHz, $\alpha$~=~$-1.3$), appears to be stable on timescales of years, and has no strong confusing off-axis sources in the field. \revone{The resulting noise level in the deep Terzan 5 image (using the same imaging parameters) is 6.6~$\mu\mathrm{Jy~beam}^{-1}$, and 300--500~$\mu\mathrm{Jy~beam}^{-1}$ in the 8~s snapshot images. This is higher than in the other two observations: variability of the calibrator (and thus an incorrrect flux scale) cannot be ruled out, however the effective noise level in the Terzan 5 images is also raised due to the presence of diffuse Galactic plane emission.}

\section{Results}

\subsection{An eclipsing MSP in Omega Centauri}

\begin{figure}
   \includegraphics[width=\columnwidth]{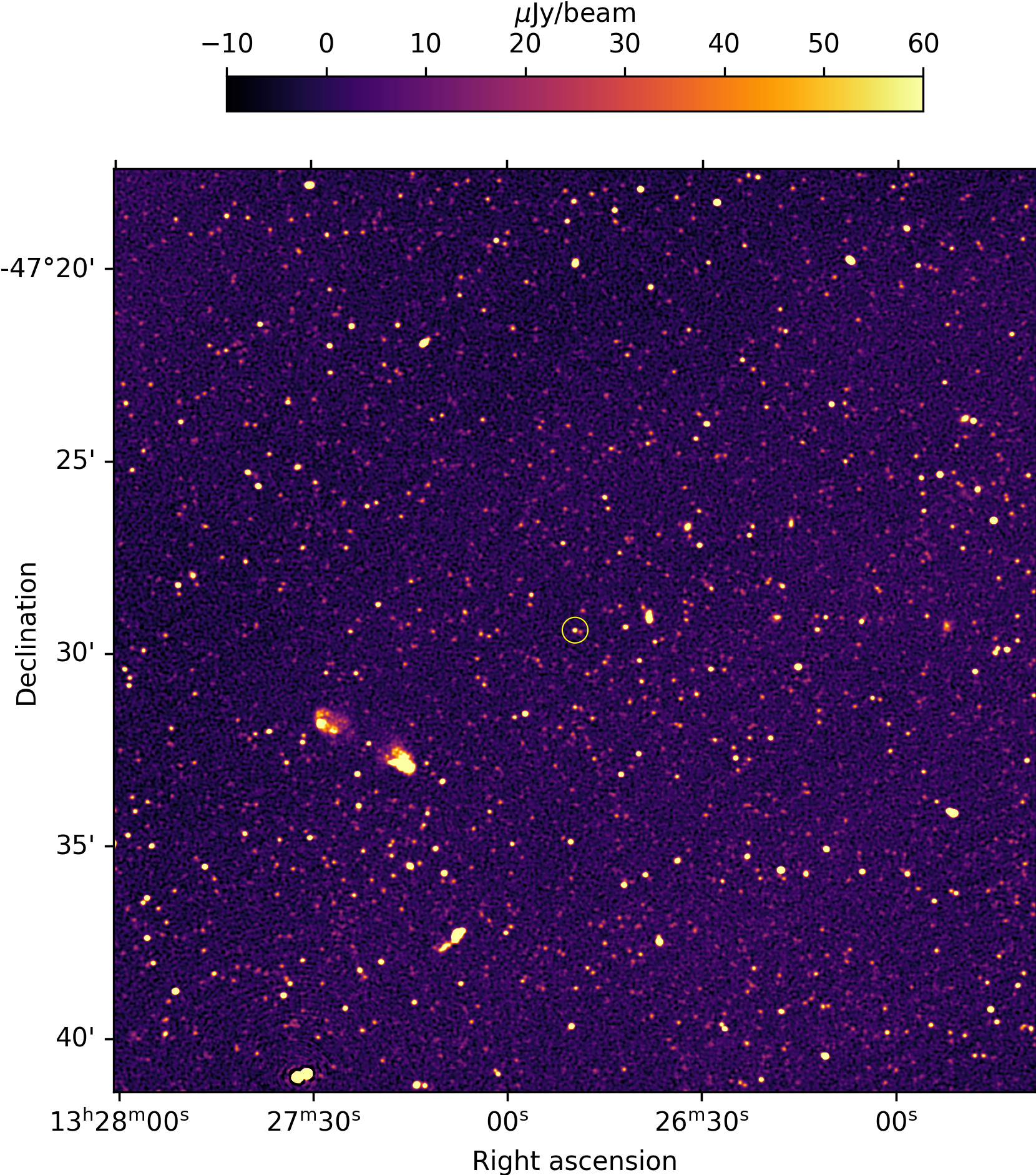}
   \caption{\label{fig:omcen}The MeerKAT image of $\omega$ Cen at 1.28 GHz, with the TRON detection of PSR J1326$-$4728B indicated. \revone{The fitted Gaussian restoring beam in this image is $5\farcs5$~$\times$~$5\farcs1$ (PA~=~87.7$^{\circ}$).}}
\end{figure}

\begin{figure} 
   \includegraphics[width=\columnwidth]{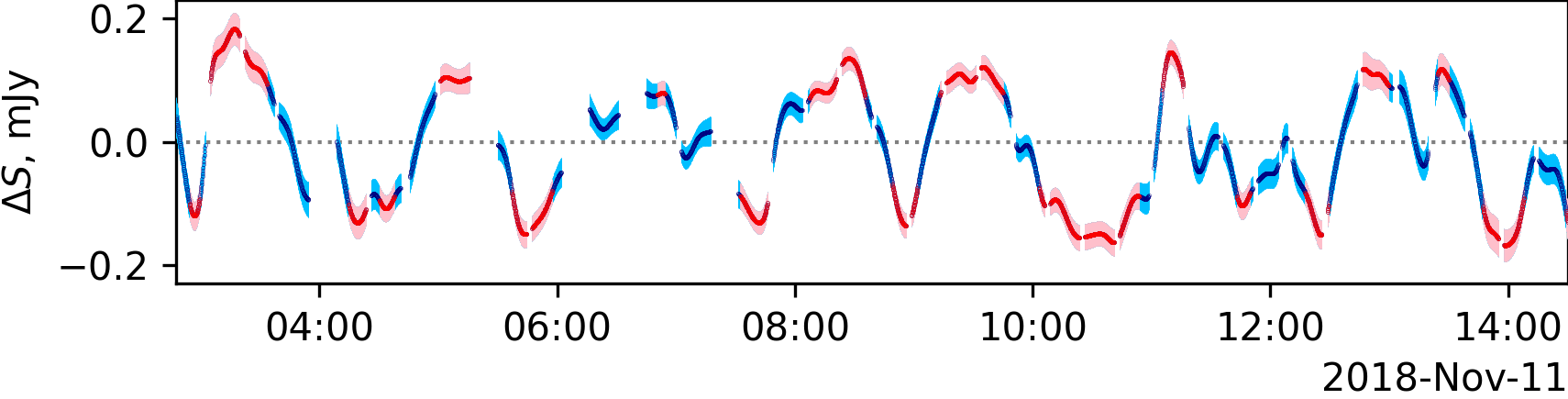}
   \caption{\label{fig:omcen-lc}Mean-subtracted lightcurve for PSR J1326$-$4728B, at 480~s smoothing. Error bars (computed as the local image rms) are plotted in light blue. Four-sigma deviations are indicated in red, with light red error bars.}
\end{figure}

Figure~\ref{fig:omcen} presents the MeerKAT L-band image of $\omega$ Cen, zoomed into the central region, highlighting the single TRON detection. This corresponds to the position of the MSP PSR J1326$-$4728B discovered by \citet{Dai2020}. A mean-subtracted lightcurve for PSR J1326$-$4728B, smoothed to 480~s, is shown in Fig.~\ref{fig:omcen-lc}.

\begin{figure}
   \includegraphics[width=\columnwidth]{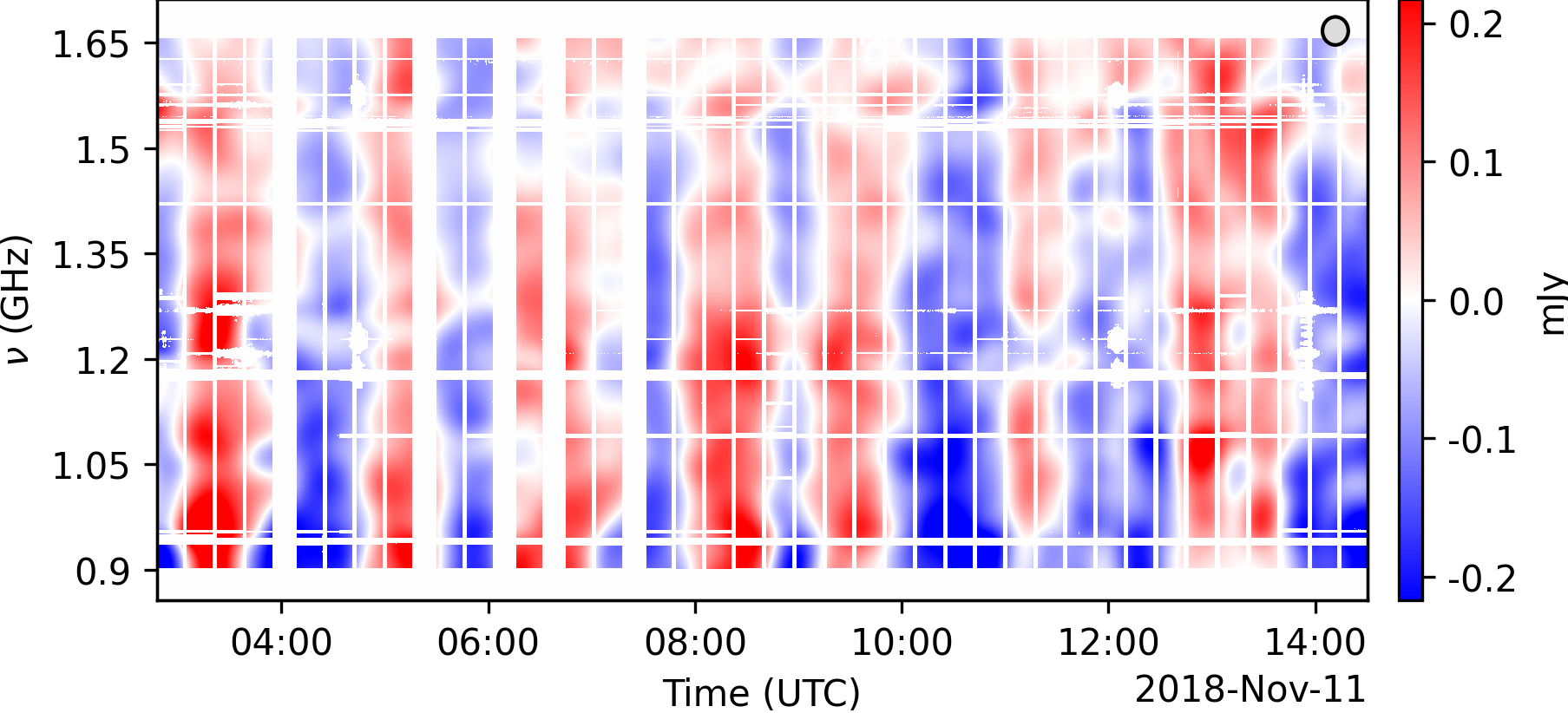}
   \caption{\label{fig:omcen-ds}L-band dynamic spectrum for PSR J1326$-$4728B, smoothed to 900~s and 40~MHz. The FWHM of the Gaussian smoothing kernel is indicated in the top right of the plot.}
\end{figure}

\revone{Omega Centauri ($\omega$ Cen, NGC 5139) is a massive and luminous globular cluster in the Milky Way, now known to host 18 MSPs, of which 7 are spider binaries \citep{Dai2020,Chen2023}. PSR J1326$-$4728B, the brightest MSP in the cluster, is the only one known to eclipse, and is established as a likely `black widow' system, or a spider with a low mass companion \citep[see also][]{Dai2023}.} 
PSR J1326$-$4728B is also associated with an X-ray source \citep{henleywillis}. \citet{Dai2020} report an orbital period of 2.15\,hr as measured from pulsar timing solutions, and describe the eclipses as ``irregular in duration''. 

In Fig.~\ref{fig:omcen-ds} we show the dynamic spectrum for PSR J1326$-$4728B generated using {\sc DynSpecMS}. The dynamic spectrum indicates the flux from the direction of the object as a function of frequency and time, with a Gaussian smoothing kernel (with a FWHM of 900~s and 40~MHz, in this case) applied to improve the signal-to-noise ratio (SNR). 

The light curve spans just over 11.5 hr, and shows multiple peaks. These range in separation from 0.75 to 1.9 hr. The largely achromatic nature of the dynamic spectrum confirms that the light curve is indeed sensitive to the eclipsing nature of PSR J1326$-$4728B, and, consistent with  \citet{Dai2020}, we observe that eclipses occur not just on the expected orbital timescale ($\sim$2.15\,hr), but also at varying orbital phases.

Some spider binary systems are known to have highly variable eclipsing environments. PSR J1740$-$5340 in NGC 6397, for example, shows irregular intensity variations, and has been found to disappear for 40\% of its long 1.35 day orbit, and yet has a well-established timing solution \citep{DAmico_2001}. The bright MSP in Terzan 5, PSR J1748-2446A, whose TRON detection is presented in Sec.~\ref{sec:Ter5}, is known to disappear for several of its 1.8\,hr orbits, where-after it can reappear to showcase regular eclipses \citep{Bilous2019, Ter5ADongzi}.  

Interestingly such variability is mostly noted in `red back' binary systems where the companion is typically a more massive main sequence star ($\gtrsim0.2$\,M$_{\odot}$), and the varying eclipses ascribed to the relativistic pulsar wind ablating the companion's surface which in turn results in a dynamic stellar outflow wind. This makes PSR J1326$-$4728B the first black widow system to exhibit irregular eclipses \citep{Dai2020}. 

\revone{Beyond the eclipsing variability, $\omega$ Cen, with a moderate cluster dispersion measure of DM~$\sim100$\,pc\,cm$^{-3}$, has also shown evidence for intensity variations due to scintillation. \citet{Dai2020} report high epoch-to-epoch variability for pulsar A. Based on this, and the lack of evidence for observed pulse broadening at 700\,MHz, they estimate the associated diffractive and refractive timescales to be $\sim10$\,min and $\sim$ few days. This confirms that the observed trends in Fig.~\ref{fig:omcen-ds} are likely dominated by the eclipsing nature of this source, although minute-timescale fluctuations could have contributions from scintillation. We also note a considerable variation between the continuum flux measurement reported by \citet{Dai2020} for pulsar B and our MeerKAT measurement ($S_{1.4}=55$ versus $133 \mu\mathrm{Jy}$). The two observations are separated by almost a year, so this variability is consistent with refractive scintillation, and that reported for pulsar A.}




\revone{TRON detected this source on timescales of 60--960~s (the latter being the longest investigated).} The variability metrics computed at 240~s (for definitions, see \citealt{47Tuc}, but note also \citealt{Heywood2024}) are $V=0.74$, $\eta=18.18$ and $\xi_{\mathrm{max}}=2.35$.

\subsection{An eclipsing MSP in Terzan 5}
\label{sec:Ter5}

\begin{figure}
   \includegraphics[width=\columnwidth]{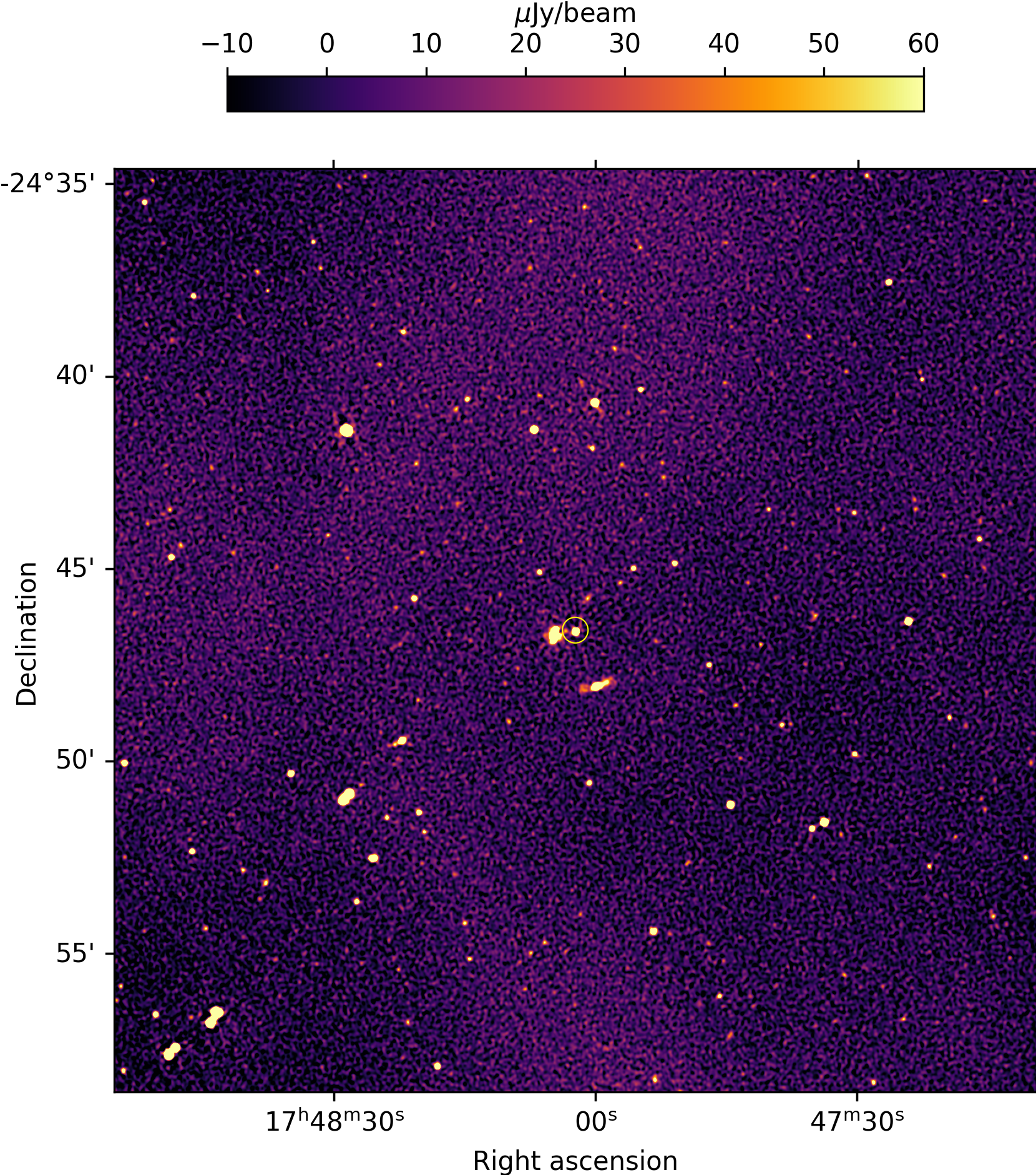}
   \caption{\label{fig:terzan5}The MeerKAT image of Terzan 5 at 1.28 GHz, with the TRON detection of Terzan 5A (J1748$-$2446A) indicated. \revone{The circular Gaussian restoring beam used in this image has a FWHM of $6\farcs1$.}}
\end{figure}

\revone{Terzan 5 is the GC with the largest known population of MSPs, with 49 confirmed to date, including 7 spiders, of which 4 are known to eclipse \citep{lyne2000,ransom2005,prager2017,andersen2018,bogdanov2020,ridolfi2021,martsen2022}. 10 of these MSPs are recent MeerKAT discoveries \citep{padmanabh2024}.} Fig.~\ref{fig:terzan5} presents the MeerKAT L-band image of Terzan 5, zoomed into the central region, highlighting the single TRON detection. The highlighted source is the MSP Terzan 5A (PSR J1748$-$2446A), the brightest known MSP in Terzan 5, and known to be an eclipsing red back system with a short orbital period of 1.8~h \citep{lyne1990} and a low-mass 0.1~M$_{\odot}$ companion star \citep{nice2000}. 

\begin{figure}
   \includegraphics[width=\columnwidth]{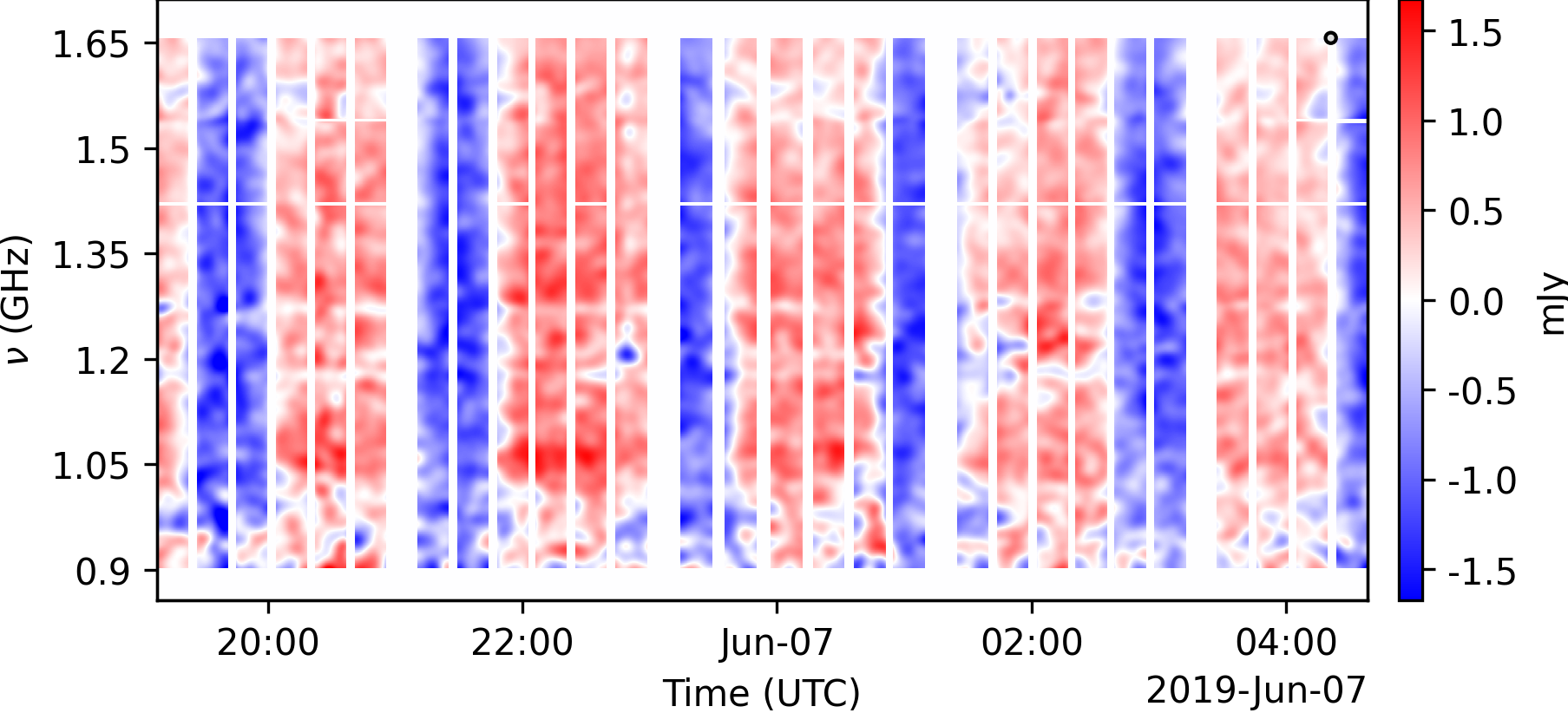}
   \caption{\label{fig:terzan5-ds}L-band dynamic spectrum for the MSP Terzan 5A, smoothed to 300~s and 15~MHz.}
\end{figure}

TRON flags up this source on timescales of 120--480~s. The variability metrics computed at 120~s are $V=5.41$, $\eta=137.27$ and $\xi_{\mathrm{max}}=1.88$. In Fig.~\ref{fig:terzan5-ds} we show the dynamic spectrum for Terzan 5A generated using {\sc DynSpecMS}, showing an eclipse-induced and mildly asymmetric duty cycle. Terzan 5 is not known to be a scintillating cluster (DM $\sim 242$ pc\, cm$^{-3}$), and we do not observe strong frequency dependent modulations across this dynamic spectrum. 

Fig.~\ref{fig:terzan5-ds} displays regular eclipsing behaviour on the anticipated orbital time scale of $\sim$1.8 \,h. 
Interestingly, however, Terzan 5A is known to exhibit eclipses of irregular durations between observing epochs, and at times even showcases additional short duration eclipses (or `mini eclipses') at varying orbital phases \citep{Bilous2019,Ter5ADongzi,SenateMSc}. The fact that our data shows regular eclipsing behaviour is in good agreement with \cite{SenateMSc}, who studied MeerKAT L-band beamformer data of this source taken a few days prior (27 May 2019). The beamformer data similarly highlights asymmetric intensity across the observed eclipses, as well as eclipse durations of $\sim$ 30 min as seen here.

Broadband pulsar data of Terzan 5A also show increasing DM values, rotation measure (RM) values and pulse profile broadening, during the eclipse ingress and egress attributed to the presence of the companion star's outflow wind \citep{Ter5ADongzi,SenateMSc}. These effects are strongly enhanced at low frequencies, and could be contributing to the increased variability in intensity observed in the lower part of the dynamic spectrum presented here. 


The remainder of the known MSPs in Terzan 5 are highly confused, visible to the east of Terzan 5A in Fig~\ref{fig:terzan5}. This motivates the need for higher angular resolution, high sensitivity observations of this GC \citep[but see also][]{urquhart2020}.

\subsection{MSPs in 47 Tucanae}

\begin{figure}
   \includegraphics[width=\columnwidth]{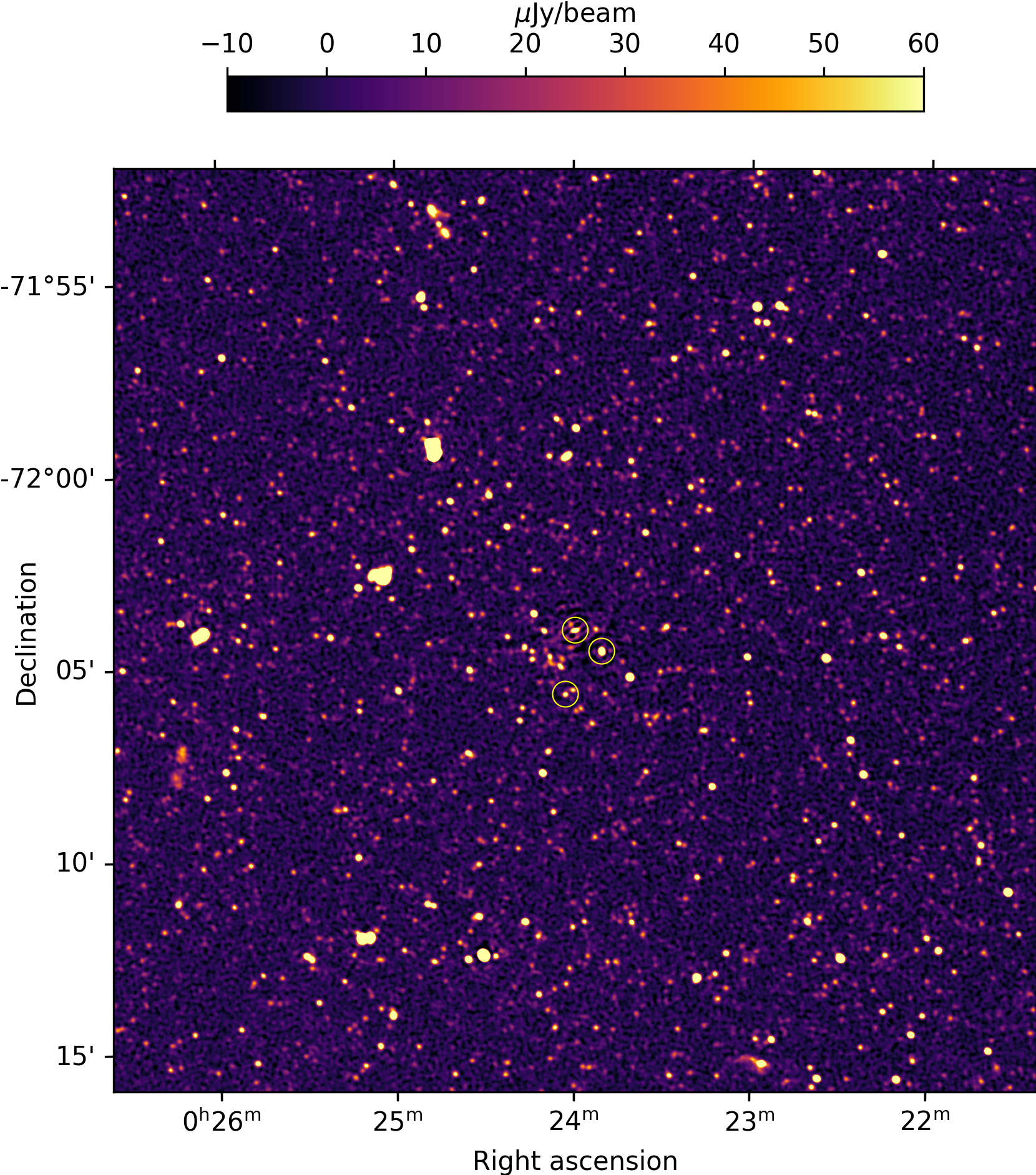}
   \caption{\label{fig:47tuc}The MeerKAT image of 47 Tuc at 1.28 GHz, with the three TRON detections indicated. The top two objects are known MSPs, while the bottom object is the MSP candidate reported by \citet{47Tuc}. The fitted Gaussian restoring beam in this image is $6\farcs9$~$\times$~$6\farcs2$ (PA~=~54.5$^{\circ}$).}
\end{figure}

\begin{table}
\begin{center}
\begin{tabular}{cccccc}
   source & $\bar{S}, \mu\mathrm{Jy}$ & $V$ & $\eta$ & $\xi_\mathrm{max}$ & $\Delta t,$ s\\
   \hline 
   J & 173 & 0.88 & 65.25 & 4.64 & 15--960 \\
   C & 526 & 0.19 & 27.02 & 4.53 & 60--960 \\ 
   $\chi$ & 101 & 0.66 & 10.47 & 3.05 & 60--480\\ 
   \hline
\end{tabular}
\end{center}
\caption{\label{tab:47tuc}Properties of detections in 47 Tuc. $\bar{S}$ is the mean flux, $V$, $\eta$ and $\xi_\mathrm{max}$ are the variability metrics \citep[see][]{47Tuc,Heywood2024} for the 480~s lightcurves, and $\Delta t$ gives the timescales on which TRON successfully detected the sources. 960~s was the longest timescale investigated.}
\end{table}

\revone{The GC 47 Tuc (NGC 104) is home to the second largest (after Terzan 5) cluster pulsar population. Of the 36 pulsars discovered in 47 Tuc, 9 are known to be spiders, with 7 of these eclipsing \citep{Manchester1991, Robinson1995,Camilo2000, Freire2001,edmonds2002,freire2003,bogdanov2005,ridolfi2016,Pan2016, Freire2017, ridolfi2021, hebbar2021, Abbate2023}.}

Figure~\ref{fig:47tuc} shows the MeerKAT L-band image of 47 Tuc, zoomed into the central region, highlighting the three TRON detections. This should be compared with Fig.~1 (top) of \citet{47Tuc}. The upper two detections  coincide with the brightest known MSPs in 47 Tuc, PSRs J0024$-$7204C and J0024$-$7204J respectively. The lower object is the proposed MSP candidate of \citet{47Tuc}, which we will label as `$\chi$'.

\begin{figure}  
   \includegraphics[width=\columnwidth]{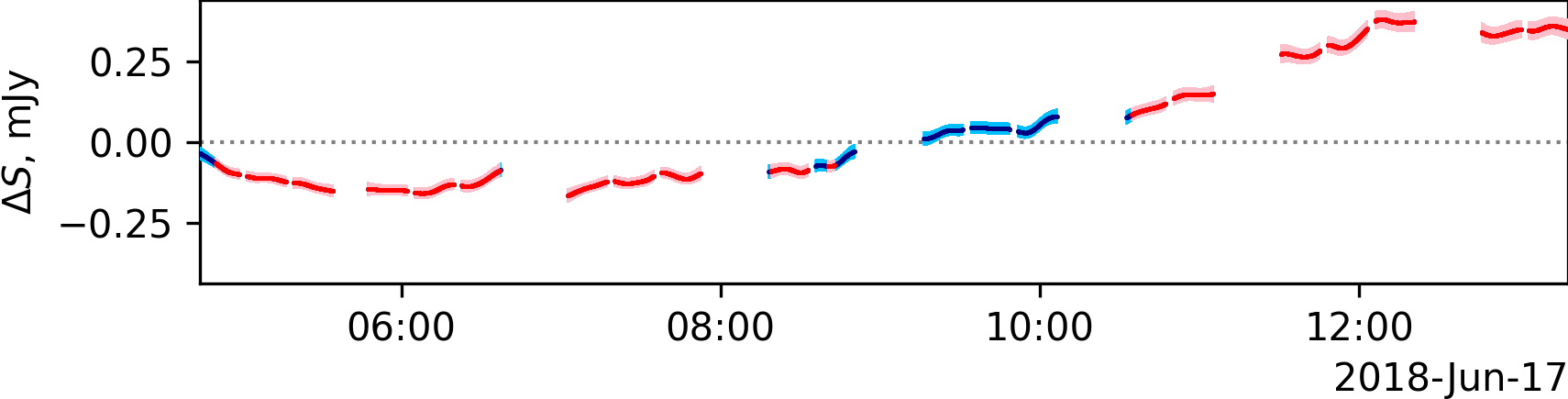}
   \includegraphics[width=\columnwidth]{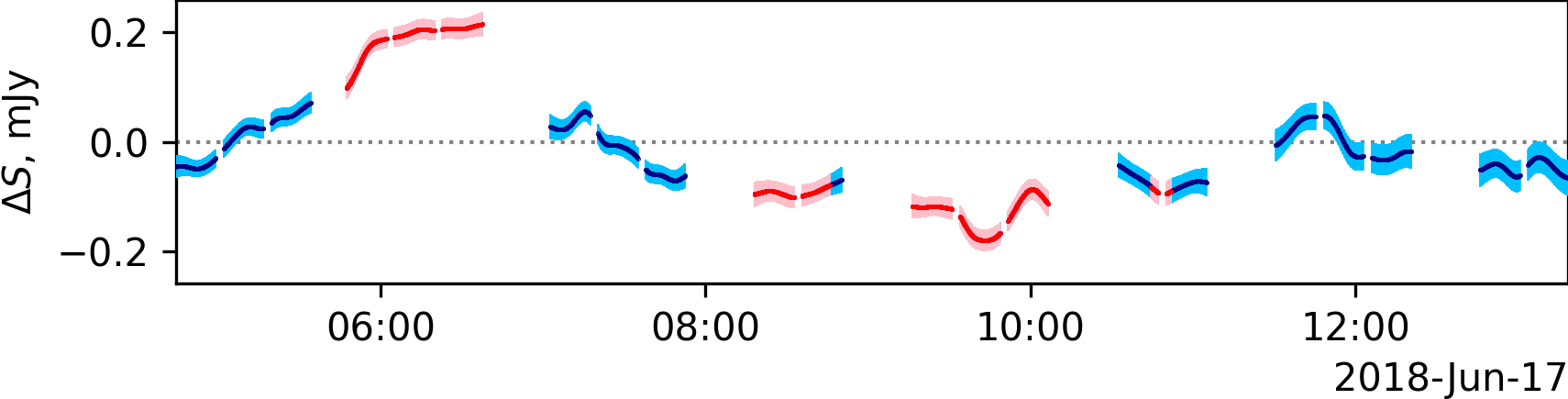} 
   \includegraphics[width=\columnwidth]{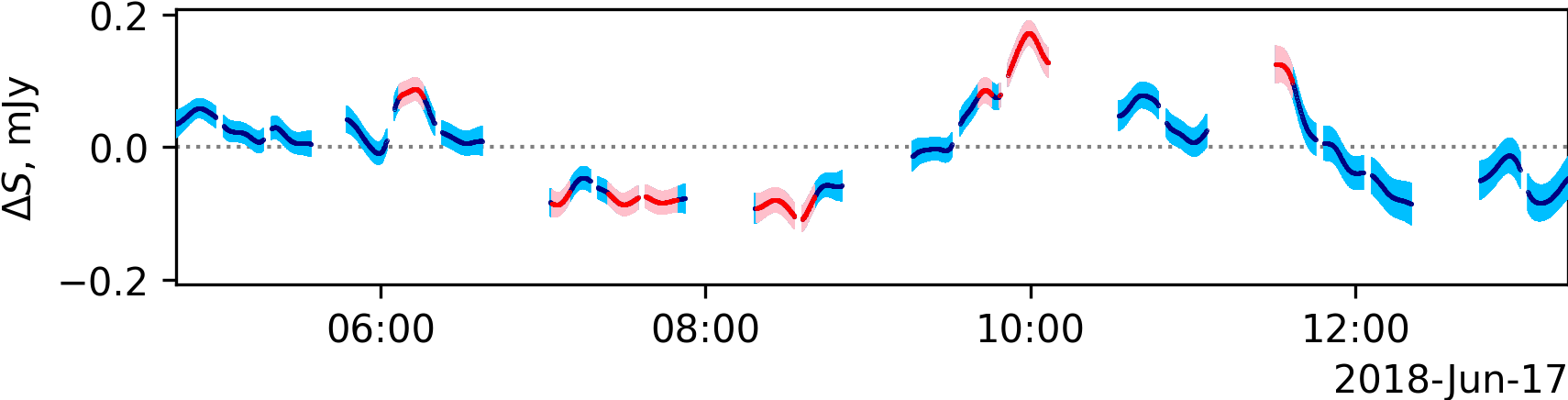}
   \caption{\label{fig:47tuc-lc}Mean-subtracted lightcurves for the three TRON detections in 47 Tuc, at 480~s smoothing. Top: MSP `J', middle: MSP `C', bottom: MSP candidate `$\chi$'. Error bars (computed as the local image rms) are plotted in light blue. Four-sigma deviations are indicated in red, with light red error bars.}
\end{figure}

Mean-subtracted light curves for these three detections, smoothed to 60s, are shown in Fig.~\ref{fig:47tuc-lc}. Table~\ref{tab:47tuc} summarizes the properties of the three sources in terms of the standard variability metrics. Note that the TRON detections are not based on these metrics per se, but rather rely on peak-finding heuristics.
Since these are intrinsically somewhat related, it is not surprising that `C', `J' and `$\chi$' show up as prominent outliers in some plots derived from the metrics by \citet[Fig. 2]{47Tuc}. Note also that \breifast\ looks for detections on multiple timescales: the timescales at which the three sources in question were detected are also indicated in the table. 

\begin{figure}
   \includegraphics[width=\columnwidth]{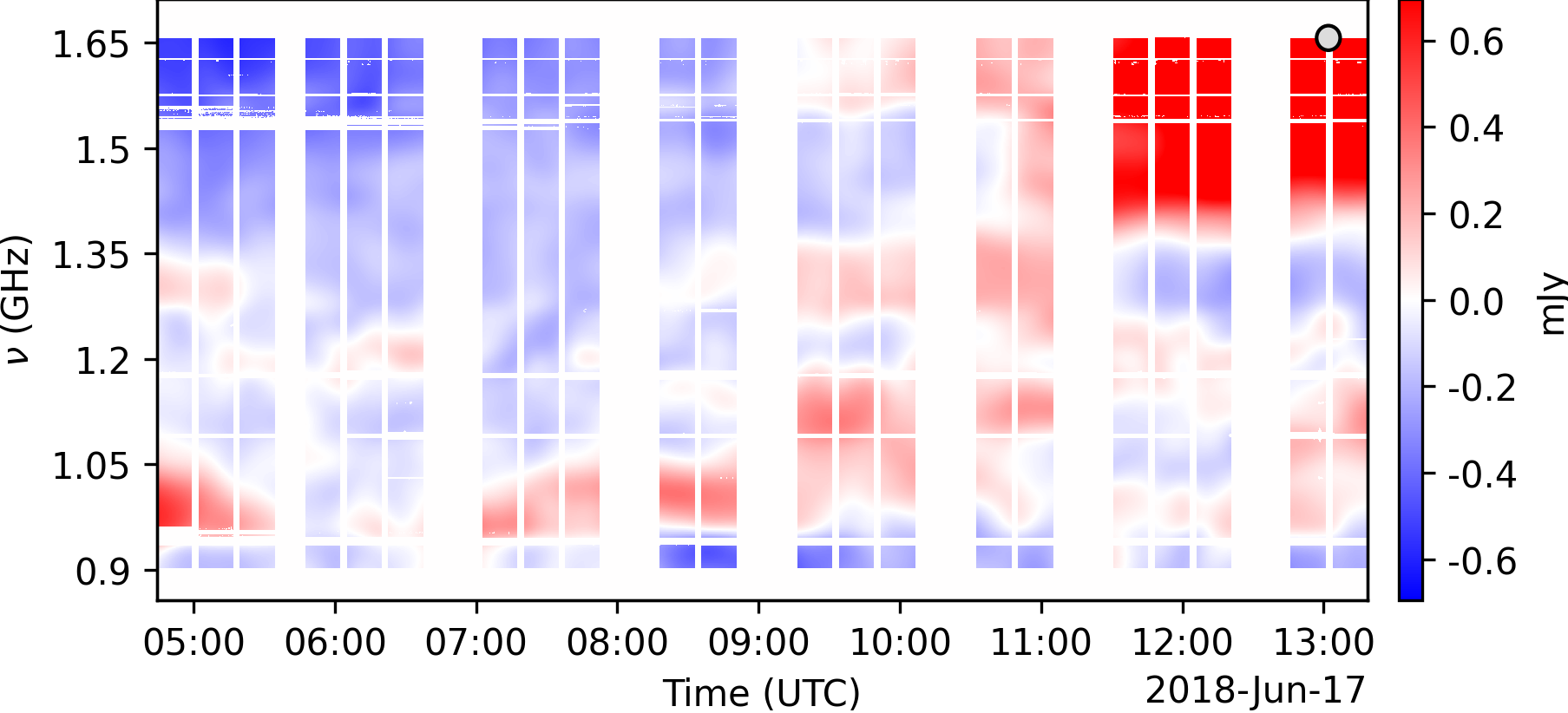}

   \vspace{1ex}
   \includegraphics[width=\columnwidth]{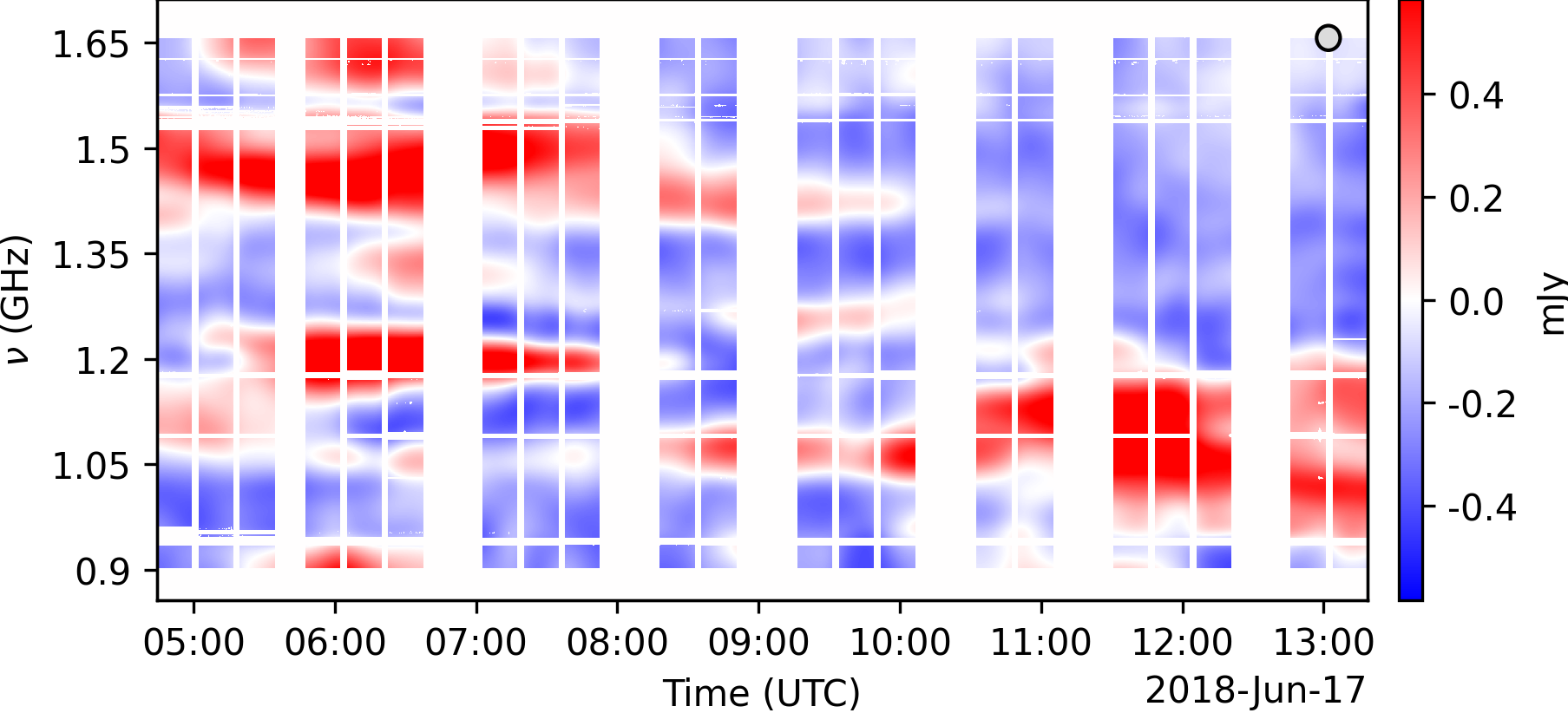}

   \vspace{1ex}
   \includegraphics[width=\columnwidth]{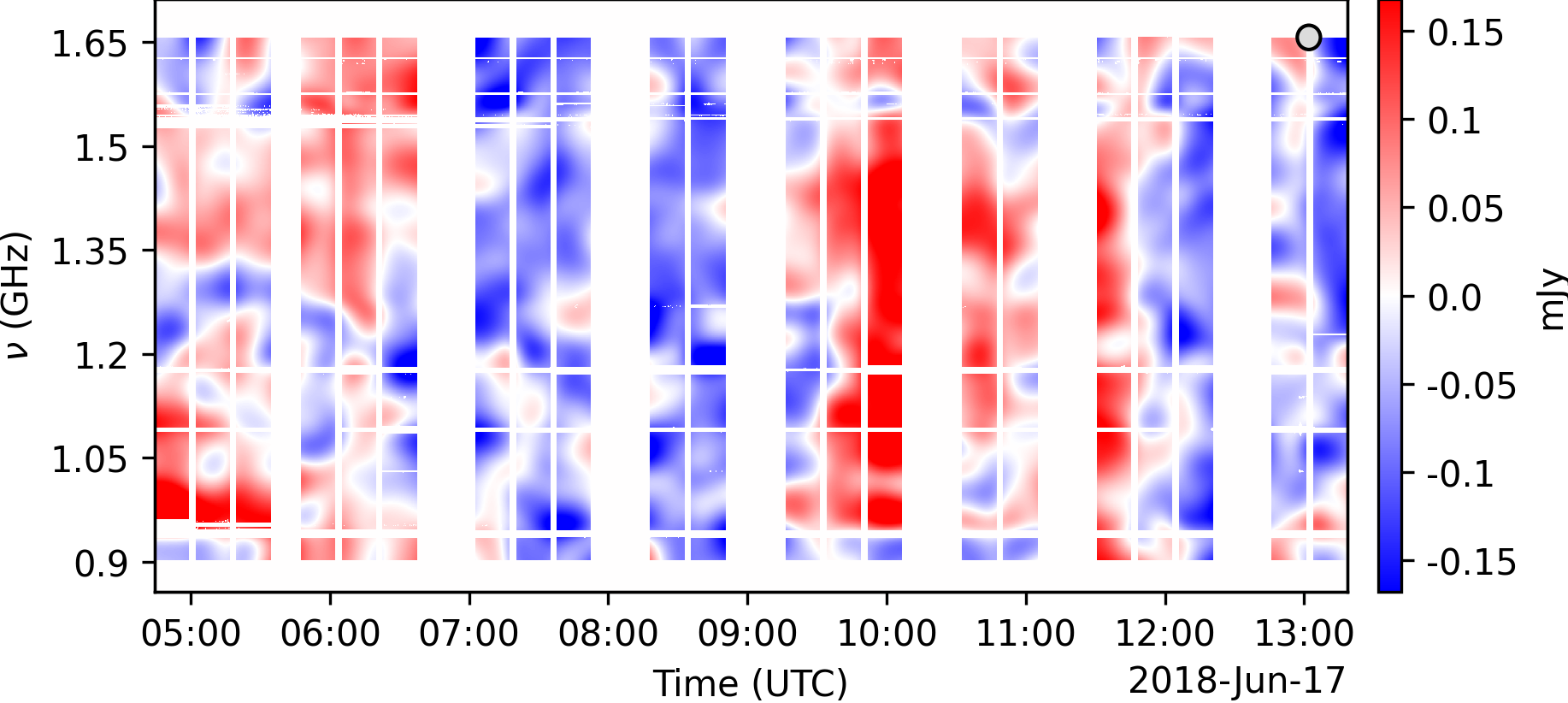}
   \caption{\label{fig:47tuc-ds}L-band dynamic spectra for the three TRON detections 
   in 47 Tuc, smoothed to 600~s and 35~MHz. Top: MSP `J', middle: MSP `C', bottom: MSP candidate `$\chi$'.}
\end{figure}

Figure~\ref{fig:47tuc-ds} shows dynamic spectra for the three sources. With a cluster DM-value of $\sim$24 pc cm$^{-3}$, 47 Tuc is known to be a scintillating cluster. We see the imprint of this scintillation in the highly chromatic variations in the dynamic spectra of especially MSPs `C' and `J'.

PSR J0024$-$7204J is a black widow pulsar which was reported to show regular eclipses at 50\,cm and 70\,cm \citep{Robinson1995}, but not at 20\,cm, or L-band \citep{Camilo2000}. We also do not find evidence for eclipses at its orbital period of 2.9\,hrs. In contrast, PSR J0024$-$7204C is an isolated pulsar, such that any variability observed cannot be due to binary motion, but must be exclusively due to scintillation.

While Fig.~\ref{fig:47tuc-ds} shows scintillation patterns for candidate `$\chi$'as well, we also note a hint of eclipsing behaviour, with two broad peaks (and corresponding troughs) separated by about 4 hours. Such a timescale would be in good agreement with a typical binary period for a spider pulsar binary system.

\section{Conclusions}

MeerKAT is proving to be a remarkable instrument for the detection of faint, medium-timescale radio transients and variables, and we can expect a systematic reprocessing of archival synthesis imaging data to yield many new discoveries. As proof of concept, we have deployed a prototype blind search pipeline, based around the TRON recipe, and applied it to a number of potentially interesting fields. \revone{In this letter, we discussed the initial detections made by TRON in three archival observations of globular clusters.}

\revone{We detect two eclipsing spider MSPs, two highly scintillating MSPs, and one candidate MSP showing hints of eclipsing. There are 12 known eclipsing spiders within these three fields; \revtwo{a follow-up study to investigate why the other spider systems are not being detected via this method will be used to refine the TRON approach.} The current version of TRON only utilizes wideband HTC images and lightcurves, which must tend to wash out chromatic effects such as scintillation. Expanding TRON to a more fine-grained search of the frequency axis would be a very promising development. In the meantime, it is already clear that eclipsing and scintillation are observational signatures that could be used to uncover more pulsars via image-plane search techniques.}

An observation of the globular cluster $\omega$ Cen yielded a detection of a known `black widow' MSP, PSR J1326$-$4728B. The nature of the variability is almost certainly intrinsic, with the lightcurve showing a clear signature of eclipsing, consistent with the results of \citet{Dai2020}. This object merits further detailed investigation. 

An observation of the globular cluster Terzan 5 yielded a detection of a known red back eclipsing pulsar, Terzan 5A. The dynamic spectrum we obtained shows eclipsing behaviour completely consistent with MeerKAT beamformer observations of the same object, which serves to confirm our techniques and interpretation of the other dynamic spectra reported on here.

The globular cluster 47 Tuc yielded blind detections of two known MSPs, and one suspected MSP already reported by \citet{47Tuc} using the same observations. The irregular nature and timescales of the variability are suggestive of propagation effects (scintillation) for the two known MSPs, while the candidate MSP shows hints of eclipsing in its dynamic spectrum. This object deserves further observational follow-up.

 
\paragraph*{Data Availability.} The raw data underlying this article is publicly available via the SARAO archive\footnote{\url{https://archive.sarao.ac.za}}, under proposal IDs SSV-20180615-FC-01, SSV-20181107-FC-02 and SSV-20190605-FC-01 for 47 Tuc, $\omega$ Cen and Terzan 5 respectively.

\paragraph*{Acknowledgements.} The MeerKAT telescope is operated by the South African Radio Astronomy Observatory, which is a facility of the National Research Foundation (NRF), an agency of the Department of Science and Innovation (DSI). The research of some of the authors is supported by the South African Research Chairs Initiative of the DSI/NRF (grant No. 81737 for OMS, JSK and VGGS, and No. 98626 for MG). We acknowledge the financial support of the Breakthrough Listen project. Breakthrough Listen is managed by the Breakthrough Initiatives, sponsored by the Breakthrough Prize Foundation. We thank Paulo Freire for maintaining a comprehensive web resource on pulsars in globular clusters which was useful for this work.

\bibliographystyle{mnras} 
\bibliography{parrot-msp} 
 
\label{lastpage}

\end{document}